\documentclass[12pt,preprint]{aastex}

\slugcomment{PASP, in press}

\shorttitle{Distance of 1FGL J1018.6--5856}
\shortauthors{Napoli et al.}

\begin{document}


\title{The Distance of the $\gamma$-ray Binary 1FGL J1018.6--5856}


\author{Vanessa J.\ Napoli\altaffilmark{1}, M.\ Virginia McSwain, Amber N.\ Marsh Boyer, and Rachael M.\ Roettenbacher\altaffilmark{2}}
\affil{Department of Physics, Lehigh University, 16 Memorial Drive E, Bethlehem, PA 18015; 
47napoli@cardinalmail.cua.edu, mcswain@lehigh.edu, anm506@lehigh.edu, rmroett@umich.edu }


\altaffiltext{1}{present address: The Catholic University of America, Physics Department, 620 Michigan Ave.\ NE, Washington, DC 20064}
\altaffiltext{2}{present address: University of Michigan, Department of Astronomy and Astrophysics, 500 Church St., Ann Arbor, MI 48109-1042}


\begin{abstract}
The recently discovered $\gamma$-ray binary 1FGL J1018.6--5856 has a proposed optical/near-infrared (OIR) counterpart 2MASS 10185560--5856459.  We present Str\"omgren photometry of this star to investigate its photometric variability and measure the reddening and distance to the system.  We find that the $\gamma$-ray binary has $E(B-V) = 1.34 \pm 0.04$ and $d = 5.4^{+4.6}_{-2.1}$ kpc.  .  While $E(B-V)$ is consistent with X-ray observations of the neutral hydrogen column density, the distance is somewhat closer than some previous authors have suggested.  
\end{abstract}





\section{Introduction}

The \textit{Fermi} Large Area Telescope source 1FGL J1018.6--5856 \citep{abdo2010} was recently discovered to have modulation of its 100 MeV--200 GeV emission with a $16.58 \pm 0.04$ d period \citep{corbet2011a}.  
\textit{Swift} observations identified a variable X-ray source with a position consistent with the location of the $\gamma$-ray source, and radio flux variations were also found using the Australia Telescope Compact Array \citep{corbet2011a}.  \citet{pavlov2011} observed the source using \textit{Chandra} and \textit{XMM-Newton}, finding absorbed power law spectra that are variable in both flux and hardness. 
The observed variability from radio to GeV energies implies that 1FGL J1018.6--5856 is a member of the elite group of ``$\gamma$-ray binaries'', which are high mass X-ray binaries that also exhibit rare $\gamma$-ray emission.  

The proposed optical and near-infrared (OIR) counterpart of the high energy source is 2MASS 10185560--5856459.  \citet{corbet2011a} find a spectral type of O6 V((f)) for the star, similar to the $\gamma$-ray binary LS 5039 \citep{mcswain2004a}.  Otherwise, knowledge of the optical star in this system is quite limited.  In this letter, we present Str\"omgren photometry of 2MASS 10185560--5856459 to investigate the optical variability of the source, the interstellar reddening $E(B-V)$, and the distance.


\section{Observations}

We observed 2MASS 10185560--5856459 using the Cerro Tololo Inter-American Observatory (CTIO) 0.9m telescope, operated by the SMARTS consortium.  We used the SITe 2048 CCD in unbinned, quad readout mode with a plate scale of 0.401\arcsec/pixel.  Observations were taken between UT dates 2011 May 20-26 and 2011 June 17-23 using the Str\"omgren $b$ and $y$ filters.  

Bias images and sky flats were taken at the start of every night.  We observed four standard stars \citep{cousins1987} at a minimum of three different air masses each night.  HD 79039, HD 80484, HD 128726, HD 157795 were used on the May run, and HD 104664, HD 105498, HD 156623, and HD 157795 were used during June.  The target was observed once each night in both filters with exposure times of 400--700 s and 200--500 s in $b$ and $y$, respectively. 

The data were reduced using standard \textit{quadproc} and \textit{cosmicray} routines in IRAF\footnote{IRAF is distributed by the National Optical Astronomy Observatory, which is operated by the Association of Universities for Research in Astronomy (AURA) under cooperative agreement with the National Science Foundation.}.  We used an aperture of 7\arcsec~to determine the instrumental magnitudes of the target and standards, and we calibrated the apparent magnitudes using the method of \citet{mcswain2004b}.  Due to poor photometric conditions on most nights, we found that only data from UT dates 2011 May 21, May 26, and June 23 were well calibrated.  Our quoted errors include both the instrumental and transformation coefficient errors as described by \citet{mcswain2004b}.  The resulting apparent magnitudes of 2MASS 10185560--5856459 are listed in Table \ref{mags}.  We find that the optical magnitudes of the star are constant within errors, but we recommend more extensive observations to investigate the variability over the complete orbital period.


\section{Reddening and Distance Measurements}

We used the mean $by$ magnitudes, with errors added in quadrature to be conservative, to determine the Str\"omgren $by$ fluxes according to \citet{gray1998}.  We converted the $JHK_s$ magnitudes of 2MASS 10185560--5856459 \citep{skrutskie2006} to fluxes using the calibration of \citet{cohen2003}.  These OIR fluxes constitute the observed spectral energy distribution (SED) of the optical star in the $\gamma$-ray binary.  

\citet{corbet2011b} present an optical spectrum of the star, which they classify as O6 V((f)).  The lack of emission in the H$\alpha$ or \ion{He}{2} $\lambda4686$ are indicative of an unevolved, main sequence star with relatively weak stellar winds.  They do observe weak emission in \ion{N}{3} $\lambda4634$, which is normal in main sequence O-type stars.  We used the calibration of \citet{martins2005} to estimate the physical properties of the star, listed in Table \ref{params}.  We used an effective temperature $T_{\rm eff} = 38,900$ K and surface gravity $\log g = 3.92$ and interpolated within the grid of Tlusty OSTAR2002 model fluxes \citep{lanz2003} to produce a model SED for 2MASS 10185560--5856459.  We binned the model SED to 50 \AA~bins to remove small scale line features.

\begin{figure}
\includegraphics[angle=90,scale=0.6]{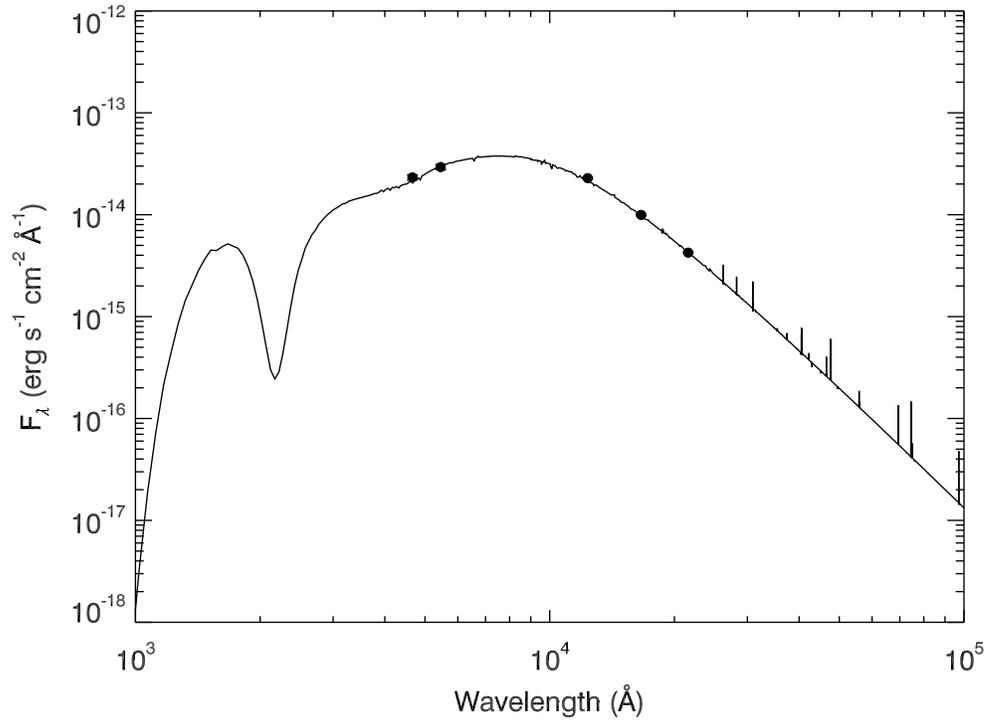} 
\caption{OIR spectral energy distribution of 2MASS 10185560--5856459.  The reddened Tlusty model SED with $T_{\rm eff} = 38,900$ K, $\log g = 3.92$, and $E(B-V) = 1.34$ is also shown, normalized to the best-fit distance $d = 5.4$ kpc.  The error bars on the observed broadband fluxes are smaller than the points shown.
\label{sed} 
}
\end{figure}

Using a grid of values for the reddening, $E(B-V)$, and a ratio of total-to-selective extinction $R = 3.1$, we applied the Galactic reddening model of \citet{fitzpatrick1999} to compare reddened model SEDs to the observed stellar fluxes.  The ratio of the observed stellar flux to the reddened model provides the stellar angular size, $\theta = R_\star/d$ \citep{gray1992}.  Using the estimated stellar radius $R_\star = 10.1 \; R_\odot$ \citep{martins2005}, the distance $d$ can thus be determined.  
We determined the best fit, $E(B-V) = 1.34^{+0.02}_{-0.03}$ and $d = 5.4 \pm 0.2$ kpc by minimizing the mean square of the deviations, rms$^2$.   The formal error in $E(B-V)$ is the offset from the best-fit value that increases the rms$^2$ by $2.7 \, \rm rms^2/N$, where $N = 5$ is the number of wavelength points within the fit region.  Our formal error in $d$ is determined by the range of allowable $E(B-V)$ values.  This reddened model SED is shown with the observed fluxes in Figure \ref{sed}.  

Another potential source of error is the lack of precisely measured $T_{\rm eff}$ and $\log g$ for 2MASS 10185560--5856459.  Based on the absence of strong emission lines, the main sequence nature of the hot star is well constrained and we allow a range of $3.80 \le \log g \le 4.25$.  $T_{\rm eff}$ is less well constrained, so we allow a 10\% error in this quantity to provide a more generous range in the allowed $E(B-V)$ and $d$.  For each $T_{\rm eff}$ and $\log g$ model, we determined $R_\star$ by interpolating between the evolutionary tracks for non-rotating stars from \citet{schaller1992}.  We repeated our distance and reddening measurements for the possible extrema of $T_{\rm eff}$ and $\log g$ which results in the more realistic error bars $E(B-V) = 1.34 \pm 0.04$ and $d = 5.4^{+4.6}_{-2.1}$ kpc.  

The observed neutral hydrogen column density, from \textit{Chandra} observations, is $nH = 0.64^{+0.19}_{-0.17} \times 10^{22}$ atoms~cm$^{-2}$ \citep{pavlov2011}.  Their \textit{XMM-Newton} observation was best fit using somewhat smaller errors for $nH$, $0.646^{+0.048}_{-0.039} \times 10^{22}$ atoms~cm$^{-2}$.  Using the relation $nH / E(B-V) = 5.8 \times 10^{21}$ atoms cm$^{-2}$ mag$^{-1}$ \citep{bohlin1978}, the \textit{Chandra} detections predict a possible range for the optical reddening of $0.81 < E(B-V) < 1.43$.  Thus our measurement of $E(B-V)$ is consistent with their results for $nH$ from \textit{Chandra}.

Our distance measurement also agrees well with the value from \citet{corbet2011b}, who estimate $5 \pm 2$ kpc.  On the other hand, our measurement is somewhat smaller than the estimate of $9 \pm 3$ kpc by \citet{pavlov2011}.  We attribute this difference to the refined value of $E(B-V)$ available due to our Str\"omgren photometry of the source.  With our improved distance, the unabsorbed luminosities of the \textit{Chandra} and \textit{XMM-Newton} observations should be revised downward to 36\% of the values provided by \citet{pavlov2011}.  


\acknowledgments

We are grateful to Arturo Gomez and the SMARTS consortium for their help scheduling and supporting these observations.  
VJN is supported by the National Science Foundation under REU site grant No.\ PHY-0849416.  RMR is supported by a NASA Harriett G.\ Jenkins Pre-doctoral Fellowship and a Sigma Xi Grant-in-Aid of Research.  This work is also supported by an institutional grant from Lehigh University.  This publication makes use of data products from the Two Micron All Sky Survey, which is a joint project of the University of Massachusetts and the Infrared Processing and Analysis Center/California Institute of Technology, funded by the National Aeronautics and Space Administration and the National Science Foundation.

{\it Facilities:} \facility{CTIO:0.9m ()}




\clearpage
\begin{deluxetable}{cccccc}
\tablecaption{Str\"omgren Photometry of 2MASS 10185560--5856459 \label{mags}}
\tablewidth{0pt}
\tablehead{
\colhead{MJD of mid} & \colhead{ } & \colhead{ } & \colhead{MJD of mid} & \colhead{ } & \colhead{ }  \\
\colhead{exposure ($b$)} & \colhead{$b$} & \colhead{$\Delta b$} & \colhead{exposure ($y$)} & \colhead{$y$} & \colhead{$\Delta y$} 
}
\startdata
55702.0557   &  13.507  &  0.052  &  55702.0595  &  12.725  &  0.045  \\
55707.1024   &  13.505  &  0.014  &  55707.1092  &  12.727  &  0.020  \\
55735.0420   &  13.518  &  0.032  &  55735.0493  &  12.851  &  0.041  \\
\enddata
\end{deluxetable}

\begin{deluxetable}{lc}
\tablecaption{Stellar Properties Assumed for 2MASS 10185560--5856459 \label{params}}
\tablewidth{0pt}
\tablehead{
\colhead{Parameter} & \colhead{Value} 
}
\startdata
Spectral Type\tablenotemark{a}~~~		& O6 V((f))	\\
Mass ($M_\odot$)\tablenotemark{b} 	&  31.0		\\
Radius ($R_\odot$)\tablenotemark{b}	&  10.1		\\
$T_{\rm eff}$ (K)\tablenotemark{b}  		& 38,900 		\\
$\log g$ (cm s$^{-1}$)\tablenotemark{b}  	& 3.92		\\
\enddata
\tablenotetext{a}{\citet{corbet2011a}}
\tablenotetext{b}{\citet{martins2005}}
\end{deluxetable}

\clearpage

\end{document}